\documentclass[a4paper]{jpconf}
\usepackage[comma,square,numbers,sort&compress]{natbib}
\usepackage{graphicx}
\usepackage{dcolumn}   
\usepackage{bm}        
\usepackage{amssymb}   
\usepackage{amsfonts}
\usepackage{graphics}
\usepackage{grffile}   
\usepackage{epsfig}
\usepackage{units}
\usepackage[usenames]{color}
\usepackage[normalem]{ulem} 
\usepackage{color}
\usepackage[utf8]{inputenc}
\usepackage[T1]{fontenc}

\usepackage{lineno}
\linenumbers
\modulolinenumbers[5]

\newcommand{\gevc}         {GeV/\ensuremath{c}}

\newcommand{\TPC}          {\rm{TPC}}

\newcommand{\VZEROA}       {\rm{VZERO-A}}

\newcommand{\pp}           {pp}

\newcommand{\Aa}           {\mbox{A--A}}

\newcommand{\pPb}          {\mbox{p--Pb}}

\renewcommand{\pt}           {\ensuremath{p_{\mathrm{T}}}{ }}

\newcommand{\snn}          {\ensuremath{\sqrt{s_{\mathrm{NN}}}}}

\newcommand{\dd}           {\ensuremath{\mathrm{d}}}

\newcommand{\Dphi}         {\ensuremath{\Delta\varphi}}
\newcommand{\Deta}         {\ensuremath{\Delta\eta}}
\newcommand{\Ntrig}        {\ensuremath{N_{\mathrm{trig}}}}
\newcommand{\Nassoc}       {\ensuremath{N_{\mathrm{assoc}}}}
\newcommand{\dNassoc}      {\ensuremath{\frac{\dd^2N_{\mathrm{assoc}}}{\dd\Deta\dd\Dphi}}}

\newcommand{\Fig}[1]       {Fig.~\ref{#1}}

\newcommand{\Ref}[1]       {Ref.~\citenum{#1}}

\newcommand{\com}[1]       {}

\graphicspath{{./img/}}

\begin{document}
\title{Two-particle correlations in p-Pb collisions at the LHC with ALICE}

\author{Leonardo Milano, on behalf of the ALICE Collaboration}

\address{CERN - Organisation européenne pour la recherche nucléaire}

\ead{Leonardo.Milano@cern.ch}

\begin{abstract}
  The double ridge structure previously observed in Pb-Pb
collisions has also been recently observed in high-multiplicity p-Pb collisions at \snn\ = 5.02 TeV
. These systems show a long-range structure (large separation in \Deta) at the near- (\Dphi~$\simeq$ 0) and away-side (\Dphi~$\simeq \pi$) of the trigger particle. 
  In order to understand the nature of this effect the two-particle correlation analysis has been extended to identified particles.
  Particles are identified up to transverse momentum \pt values of 4 \gevc\ using the energy loss signal in the Time Projection Chamber detector, complemented with the information from the Time of Flight detector. This measurement casts a new light on the potential collective (i.e. hydrodynamic) behaviour of particle production in p-Pb collisions.
\end{abstract}

\section{Introduction}
The study of particle correlations is a powerful tool to probe the mechanism of particle production in collisions of hadrons and nuclei at high beam energy.
This is achieved  by measuring the distributions of relative angles $\Dphi$ and $\Deta$, where $\Dphi$ and $\Deta$ are the differences in azimuthal angle~$\varphi$ and pseudorapidity~$\eta$ between two particles.
In small systems, such as minimum-bias proton--proton~(\pp) collisions, the correlation at ($\Dphi \approx 0$, $\Deta \approx 0$) is dominated by the ``near-side'' jet peak, and at $\Dphi \approx \pi$ by the recoil or ``away-side'' structure due to particles originating from jet fragmentation~\cite{Wang:1992db}.
Additional ridge-like structures, which persist over a long range in $\Deta$, emerge in nucleus--nucleus~(\Aa) collisions in addition to the jet-related correlations~\cite{Aamodt:2011by,Chatrchyan:2012wg,ATLAS:2012at} .
Recent measurements in \pPb\ collisions employed a procedure for removing the jet contribution by subtracting the correlations extracted from low-multiplicity events, revealing essentially the same long-range structures on the near and away side in high-multiplicity events~\cite{alice_pa_ridge,atlasridge}.
These ridge structures have been attributed to mechanisms that involve initial-state effects, such as gluon saturation~\cite{Dusling:2013oia} and colour connections forming along the longitudinal direction~\cite{Arbuzov:2011yr}, and final-state effects, such as parton-induced interactions~\cite{Alderweireldt:2012kt}, and collective effects developing in a high-density system possibly formed in these collisions~\cite{Bozek:2013uha}.
To further characterize this effect in \pPb\ collisions at $\snn=$ \unit[5.02]{TeV}, these ridge structures are studied via a Fourier decomposition~\cite{Voloshin:1994mz} and the $v_2$ of pions, kaons and
protons\footnote{Pions, kaons and protons, as well as the symbols $\pi$, K and p, refer to the sum of particles and antiparticles.} has been measured.
\section{Analysis}
\begin{figure}[ht!f]
  \centering
  \includegraphics[width=.8\textwidth]{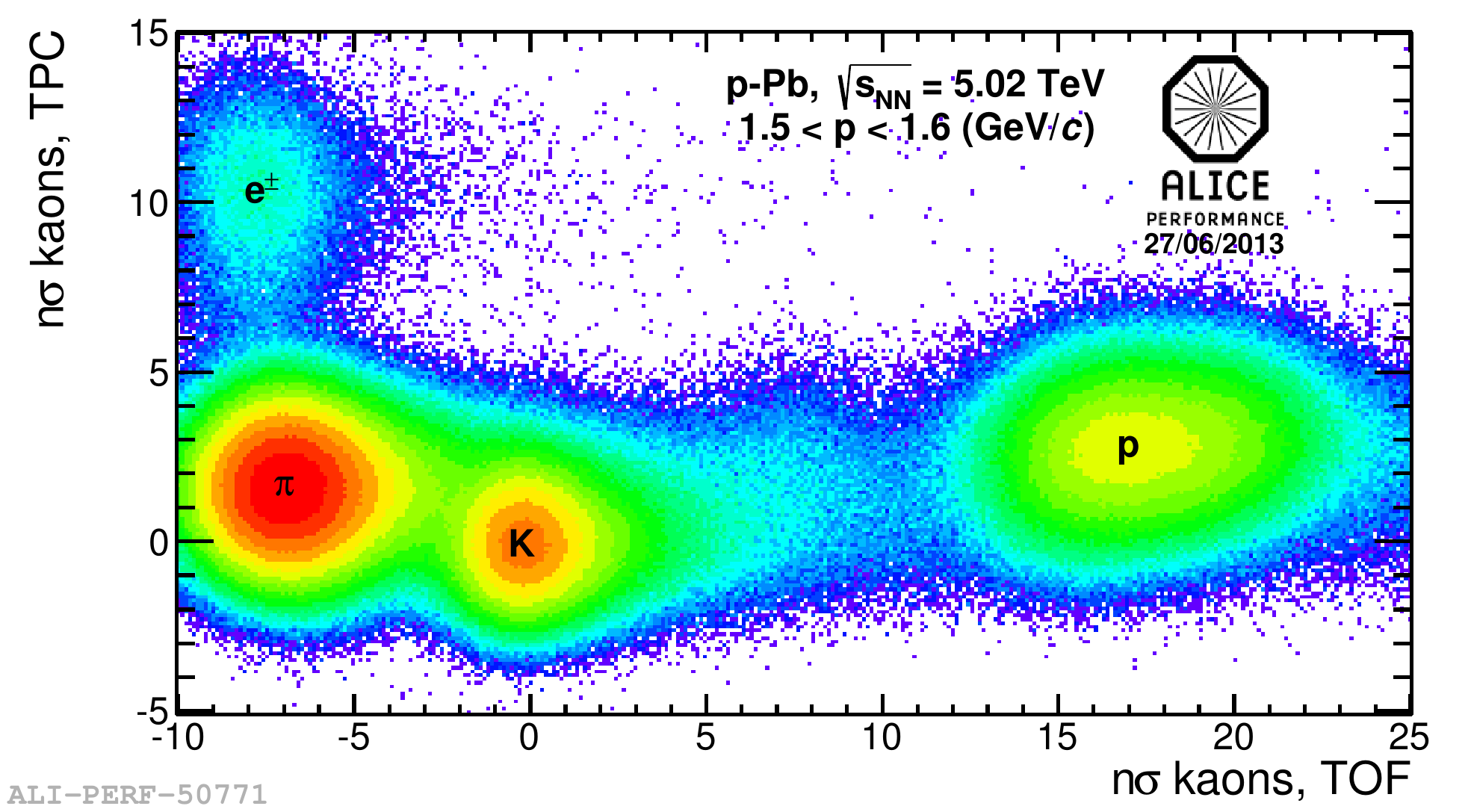}
  \caption{\label{fig:nsigma}
    $N_{\sigma, \rm TPC}$ versus $N_{\sigma, \rm TOF}$ for tracks with $1.5<p<$~\unit[1.6]{\gevc} in the kaon mass hypothesis.}
\end{figure}

A detailed description of the ALICE detector and the event and track selection can be found in \Ref{Aamodt:2008zz,alice_pa_ridge}.
Events are classified in four classes defined as fractions of the analyzed event sample, based on the charge deposition in the \VZEROA\ detector, and denoted ``0--20\%'', ``20--40\%'', ``40--60\%'', ``60--100\%'' from the highest to the lowest multiplicity.
Particle identification is based on the difference (expressed in units of the resolution $\sigma$ - $N_{\sigma}$) between the measured and the expected signal for $\pi$, K, or p in the Time Projection Chamber~(\TPC) and the Time Of Flight detector~(TOF) detectors. The $N_{\sigma, \rm TPC}$ versus $N_{\sigma, \rm TOF}$ correlation is reported in \Fig{fig:nsigma} for tracks with momentum ($p$) $1.5<p<$~\unit[1.6]{\gevc}, when the kaon mass is assumed.
For a given species, particles are selected with a circular cut defined as $\sqrt{N_{\sigma, \rm TPC}^2 + N_{\sigma, \rm TOF}^2} < 3$.
In the region where the areas of two species overlap, the identity corresponding to the smaller distance is assigned.
Contamination from misidentified particles is significant only for K above \unit[1.5]{\gevc} and is less than 15\%.

This analysis uses unidentified charged tracks as trigger particles and combines them with either unidentified charged hadrons or with $\pi$, K and p as associated particles (denoted $h-h$, $h-\pi$, $h-{\rm K}$ and $h-{\rm p}$, respectively).
The correlation is expressed in terms of the associated yield per trigger particle where both particles are from the same transverse momentum $\pt$ interval in a fiducial region of $|\eta|<0.8$:
\begin{equation}
  \frac{1}{\Ntrig} \dNassoc = \frac{S(\Deta,\Dphi)}{B(\Deta,\Dphi)} \label{pertriggeryield}
\end{equation}
where $\Ntrig$ is the total number of trigger particles in the event class and $\pt$ interval.
The signal distribution $S(\Deta,\Dphi) = 1/\Ntrig\ \dd^2N_{\rm same}/\dd\Deta\dd\Dphi$ is the associated yield per trigger particle for particle pairs from the same event.
The background distribution $B(\Deta,\Dphi) = \alpha\ \dd^2N_{\rm  mixed}/\dd\Deta\dd\Dphi$ is constructed by correlating the trigger particles in one event with the associated particles from other events of the same event class and within the same \unit[2]{cm}-wide $z_{\rm vtx}$ interval and corrects for pair acceptance and pair efficiency.

\section{Results}
\begin{figure}[ht!f]
  \centering
  \includegraphics[width=.95\textwidth]{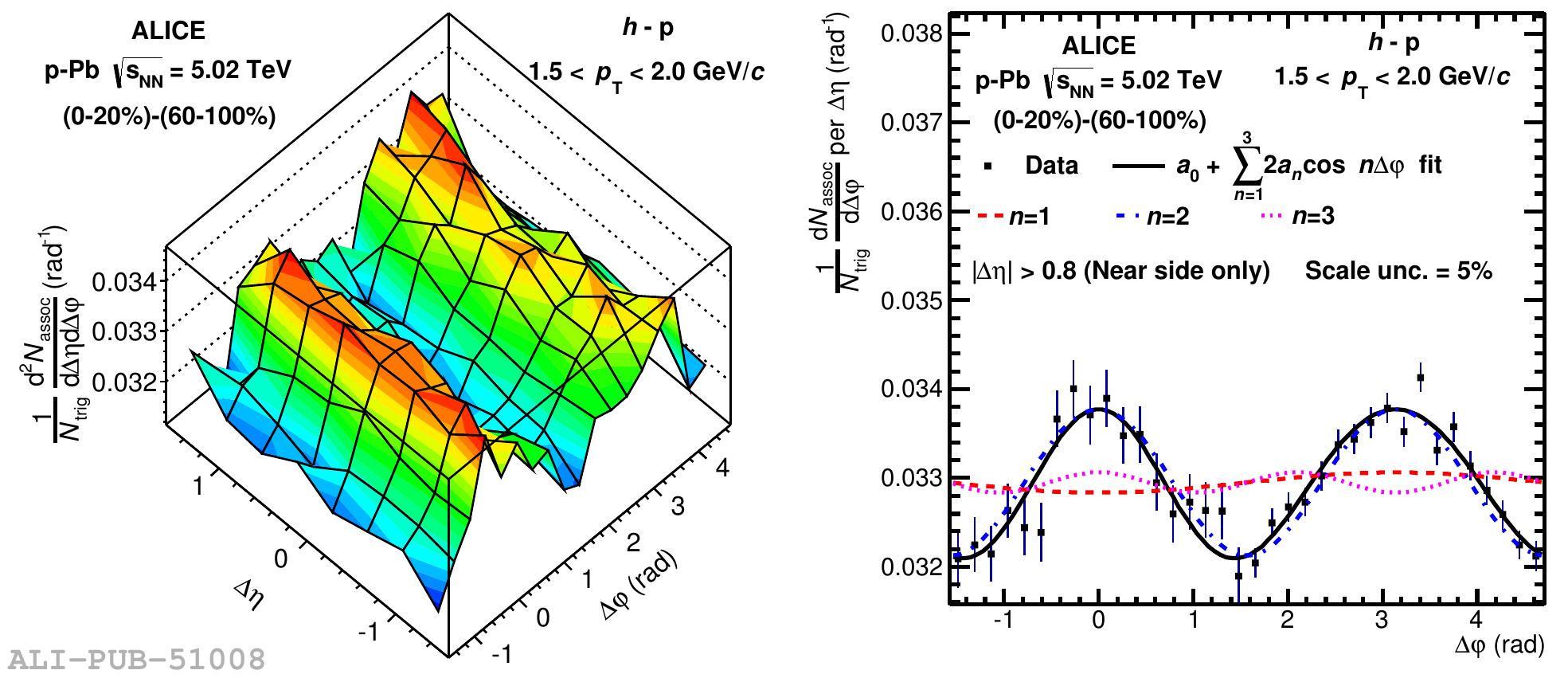}
  \caption{\label{fig:pertriggeryields_subtracted}
    Left panel: associated yield per trigger particle as a function of $\Dphi$ and $\Deta$ for $h-{\rm p}$ correlations for $1.5<\pt<$~\unit[2]{\gevc} for the 0--20\% event class where the corresponding correlation from the 60--100\% event class has been subtracted. Right panel: projection of the left panel to $\Dphi$ averaged over $0.8 < |\Deta| < 1.6$ on the near side and $|\Deta| < 1.6$ on the away side. The figure contains only statistical uncertainty. Systematic uncertainties are mostly correlated and are less than 5\%.
  }
\end{figure}

The per-trigger yield of the 60--100\% event class is subtracted from that in the 0--20\% event class in order to reduce the jet contribution as in \Ref{alice_pa_ridge}.
In the left panel of Fig.~\ref{fig:pertriggeryields_subtracted} the resulting $h-{\rm p}$ correlation for $1.5<\pt<$~\unit[2]{\gevc} is shown.
Fourier coefficients can be extracted from the $\Dphi$ projection of the per-trigger yield by a fit with:
\begin{equation}
  \frac{1}{\Ntrig} \frac{\dd \Nassoc}{\dd\Dphi} = a_0 + 2\,a_1 \cos \Dphi + 2\,a_2 \cos 2\Dphi + 2\,a_3 \cos 3\Dphi.
  \label{fitfunction1}
\end{equation}
The projection is averaged over $0.8 < |\Deta| < 1.6$ on the near side and $|\Deta| < 1.6$ on the away side.
From the relative modulations $V_{n\Delta}^{h-i}\{{\rm 2PC, sub}\} = a_n^{h-i} / (a_0^{h-i}+b)$, where $a_{n}^{h-i}$ is the $a_{n}$ extracted from $h-i$ correlations and $b$ is the combinatorial baseline of the lower-multiplicity class which has been subtracted ($b$ is determined on the near side within $1.2 < |\Deta| < 1.6$), the $v_n\{{\rm 2PC,sub}\}$ coefficient of order $n$ for a particle species $i$ (out of $h$, $\pi$, K, p) are then defined as:
\begin{eqnarray}
 v_n\{{\rm 2PC,sub}\} = \sqrt{V_{n\Delta}^{h-h}} \hspace{2cm} v_n\{{\rm 2PC,sub}\} = V_{n\Delta}^{h-i} / \sqrt{V_{n\Delta}^{h-h}}. \label{vn}
\end{eqnarray}

\begin{figure}[ht!f]
  \centering
  \includegraphics[width=0.9\textwidth]{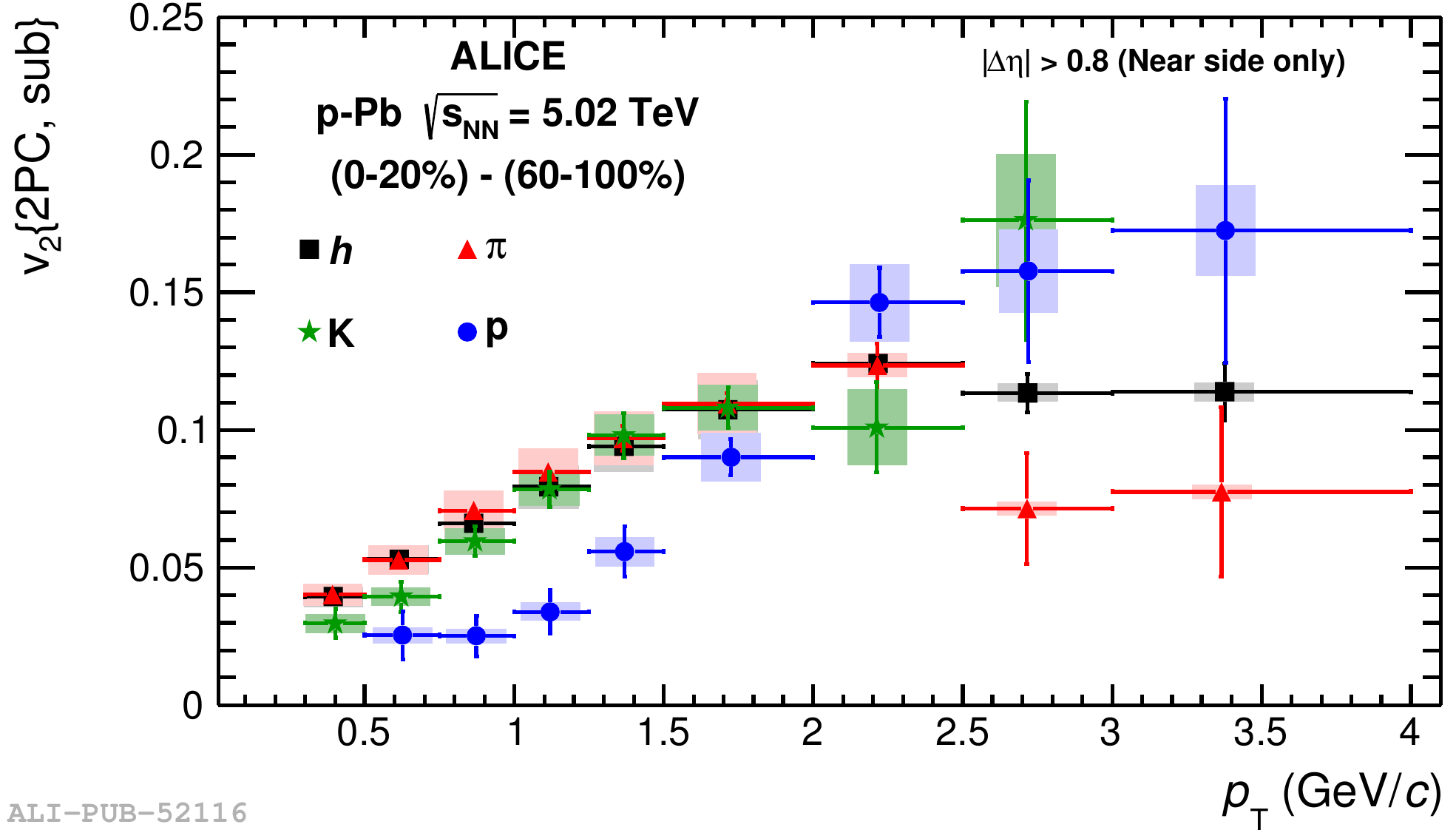}
  \caption{\label{fig:v2_subtracted}
    The Fourier coefficient $v_2\{{\rm 2PC,sub}\}$ for hadrons (black squares), pions (red triangles), kaons (green stars) and protons (blue circles) as a function of $\pt$ from the correlation in the 0--20\% multiplicity class after subtraction of the correlation from the 60--100\% multiplicity class. The data are plotted at the average-$\pt$ for each considered $\pt$ interval and particle species under study. Error bars show statistical uncertainties while shaded areas denote systematic uncertainties.
  }
\end{figure}
Figure~\ref{fig:v2_subtracted} shows the extracted $v_2\{{\rm 2PC,sub}\}$ coefficients for $h$, $\pi$, K and p as a function of $\pt$.
The coefficient $v_2^{\rm p}$ is significantly lower than $v_2^\pi$ for $0.5 < \pt <$~\unit[1.5]{\gevc}, and larger than $v_2^\pi$ for $\pt >$~\unit[2.5]{\gevc}. The crossing occurs at $\pt \approx$ \unit[2]{\gevc}. The coefficient $v_2^{\rm K}$ is consistent with $v_2^\pi$ above \unit[1]{\gevc}; below \unit[1]{\gevc} there is a hint that $v_2^{\rm K}$ is lower than $v_2^\pi$.
The mass ordering and crossing is qualitatively similar to observations in nucleus--nucleus collisions \cite{Abelev:2012di}.
\section{Summary}
The Fourier coefficient $v_2$ of these double-ridge structures exhibits a dependence on $\pt$ that is reminiscent of the one observed in collectivity-dominated Pb--Pb collisions at the LHC.
These observations and their qualitative similarity to measurements in \Aa\ collisions \cite{Abelev:2012di} are rather intriguing.  Furthermore, a mass ordering at low transverse momenta can be described by hydrodynamic model calculations \cite{Werner:2013ipa,Bozek:2013ska}. Their theoretical interpretation is promising to give further insight into the unexpected phenomena observed in \pPb\ collisions at the LHC.
\bibliographystyle{iopart-num}
\bibliography{biblio}{}

\end{document}